\documentclass{iaus}
\usepackage{graphicx}

\title[Secular Evolution] 
{Galaxy Dynamics: Secular Evolution and Accretion}

\author[F. Combes]   
{Francoise Combes}

\affiliation{Observatoire de Paris, LERMA and CNRS,
61 Av. de l'Observatoire, F-75014 Paris, France \\ email: {\tt francoise.combes@obspm.fr} }

\pubyear{2010}
\volume{271}  
\pagerange{111--118}
\jname{Astrophysical Dynamics: From Stars to Galaxies}
\editors{N. Brummell \& A.S. Brun, eds.}
\begin{document}

\maketitle

\begin{abstract}
Recent results are reviewed on galaxy dynamics, bar evolution,
destruction and re-formation, cold gas accretion, gas radial flows and AGN
fueling, minor mergers. Some problems of galaxy evolution are
discussed in particular, exchange of angular momentum,
radial migration through resonant scattering, and consequences on abundance
gradients, the frequency of bulgeless galaxies, and the relative role of
secular evolution and hierarchical formation.
\end{abstract}

\firstsection 
\section{Disk formation and "viscosity"}
  There are at least three main scenarios invoked for galaxy formation:
the first one, inspired by Eggen et al (1962) is the monolithic collapse, where the 
initial gas clouds collapse gravitationally while forming quickly stars 
before the end of their collapse, 
so that they end up with a stellar spheroid. Disks must then form later on, through
slower gas accretion. The second is the hierarchical scenario, where the first systems
to form are disks (the star formation time-scale being longer than the collapse time).
Then the interaction and merger of two disks lead to the formation of a spheroid.
The third scenario, developped in this talk, considers that disks form first, 
as in the second scenario.
However, disks then may evolve without any merger with other galaxies. 
Their own internal evolution could also produce a spheroid at the center, 
through secular evolution. This scenario assumes external gas accretion from 
filaments of the cosmic web (e.g. Kormendy \& Kennicutt 2004).

Spontaneously, a disk evolves through its gravitational instability, producing
non-axisymmetric features or waves, that will transfer angular momentum. This
is equivalent to an effective kinematic viscosity (Lin \& Pringle 1987a).
The stability of the disk is ensured at small scales by the equivalent
pressure, due to the disordered motions, or velocity dispersions. All
scales smaller than the Jeans length $\lambda_J=\sigma t_{ff} =
 \sigma/(2\pi G\rho)^{1/2}$ are stabilised, where $\sigma$ is the velocity dispersion,
and $\rho$ the density. At large scales, the disk is stabilised by the differential
rotation, and the subsequent shear. All scales larger than the critical $L_{crit}$
are stable, where $L_{crit} \sim G \Sigma / \kappa^2$, with $\Sigma$ the disk surface
density, and $\kappa$ the epicyclic frequency. Scales larger than $\lambda_J$
and lower than $L_{crit}$ remain unstable, unless the velocity dispersion
is increased until  $\lambda_J=L_{crit}$: this condition is the 
Toomre criterion (Toomre 1964). The parameter $Q^2 = \lambda_J/L_{crit}$ must
be larger than 1 for stability. From hydrostatic equilibrium in the z-direction,
the disk thickness must be $h = h_r \, {\rm min} (1,Q)$, with $h_r = \sigma/\Omega$.

When instabilities occur, they transfer momentum on scale $L_{crit}$, 
with time scale $\Omega^{-1 }$ (or dynamical time-scale); therefore
a prescription for effective viscosity  is
$\nu \sim L_{crit}^2/\Omega^{-1} \sim Q^{-2} h_r^2 \Omega$. 
In addition, it can be shown that, when the viscous time-scale is of the
same order as the star-formation time-scale,
i.e. $t_\nu \sim t_*$, then an exponential disk is formed (Lin \& Pringle 1987b).

\section{Bars and gas flows}

Bars are the most frequent non-axisymmetries developped in galaxy disks.
75\% of spiral galaxies are barred, when viewed in the near-infrared.
 They form easily in numerical simulations, even when $Q > 1$.
Bars are very efficient to trasnfer angular momentum.
They are robust and generate long-lived gravity torques,
acting on the gas, to concentrate mass towards the center.
The amount of gas driven towards the center can be
quantified by observations. It can be shown that the sign
of the torques change at each Lindblad resonance, in particular
they are negative (with respect to the sense of rotation) 
inside corotation (CR),
and positive outside. The amplitude of the torques depends on the
bar strength, and on the phase shift between the gas response and the
stellar potential. Inside CR, the gas leads the stars, and this 
corresponds to the characteristic morphology of dust lanes
leading the bar, commonly observed in barred galaxies.

  From the near-infrared image, representing the old stellar component,
i.e. most of the mass in the visible disk, it is possible to deduce 
the gravitational potential
of the galaxy. From the gas map observed either through HI or CO lines, 
it is then possible to compute the average torque exerted on the gas
at each radius, assuming a stationary state at least for one rotation.
These computations have shown that typically in a strongly barred
galaxy, the gas may lose 30\% of its angular momentum at each rotation
(Garcia-Burillo et al 2005).
 The stellar bar receives the angular momentum lost by the gas,
and therefore weakens: the bar is a wave with negative momentum,
with most the orbits sustaining the bar with high eccentricity.
The absorption of angular momentum makes the orbits rounder and the bar
weaker.  This can lead to the destruction of the bar 
(e.g. Bournaud \& Combes 2002).

The gas driven towards the center could fuel the supermassive black
hole, present in every galaxy with bulge, and trigger nuclear activity (AGN).
It is however difficult to find direct evidence of correlation
between AGN and bars, since the evolution time-scales are widely different:
the dynamical time-scale for the gas to flow to the center, and to weaken
the bar is a few 10$^8$ yr, while the AGN active phase is a few  10$^7$ yr.
  To account for the large frequency of bars,  in spite of their easy 
weakening and destruction, another mechanism has to re-form bars:
this is done through external gas accretion, which replenishes the disk,
and makes it unstable again to bar formation.
 First gas is stalled outside the outer resonance by the positive torques,
when the bar is strong. When the bas is weakened, the external gas can then 
enter and fuel the disk, intermittently.

There is therefore a self-regulated cycle of bar
formation and destruction: the bar forms in an unstable cold disk,
rich in gas. The strong bar produces gas inflow, which
weakens or destroys the bar. Gas accretion can then enter and 
re-juvenate the disk, and a new bar form.
A few percent in mass of gas infall is enough to transform a bar in a lens
(Friedli et al 1994, Berentzen et al 1998, Bournaud et al 2002, 2005).
 External gas accretion is essential to drive the secular 
evolution of galaxy disks, and 
to maintain spirals and bars frequent enough in galaxies.
The observed bar frequency can be used to quantify the accretion rate
(Block et al 2002). The required accretion rate corresponds to
the baryonic galaxy mass doubling in 7 Gyrs.
Cosmological simulations have recently revealed the importance of
cold gas inflow in filaments. The predicted  
rate of gas accretion is similar to what is required to maintain 
the observed bar frequency (Dekel \& Birnboim 2006).

\begin{figure}[b]
\begin{center}
 \includegraphics[width=8.2cm]{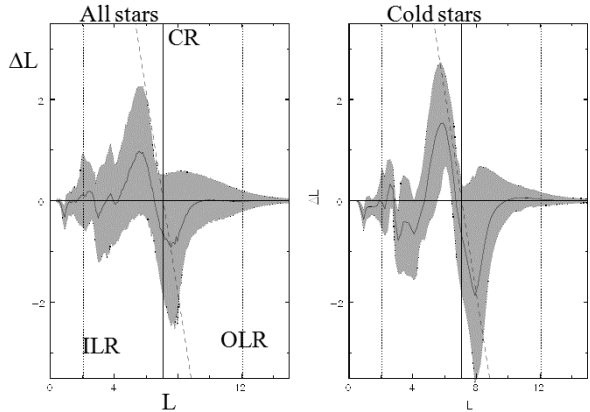} 
 \caption{Exchange of angular momentum $\Delta$L, as a function of L,
for all stars (left) and only stars with small epicycles (right).  
The locations of Lindblad resonances ILR, CR and OLR are indicated by vertical lines. 
The resonance scattering 
occurs mainly at corotation, from Sellwood \& Binney (2002).}
   \label{fig1}
\end{center}
\end{figure}

A consequence of angular momentum redistribution over
the disk by the bar, is the formation of exponential disks,
with radial breaks, and different slopes in the outer parts.
 The break is expected at the outer Lindblad resonance
(Pfenniger \& Friedli 1991). When several patterns develop
in the disk, there could be several breaks at differnt radii.
Alternatively, breaks may also be the consequence of 
gas accumulation, due to external accretion, together
with a threshold of star formation in gas surface density
(Roskar et al 2008). The galaxy disk then form inside out,
and the break moves radially outwards.
This is confirmed through numerical simulations, which 
show how the successive spirals and bars transfer some
of the inner stars to the outer parts. Age and abundance 
gradients can then change suddenly in the outer parts,
after the break limiting the location of new star formation.
This can explain the frequent observations of these surprising 
reversals (for instance after a radius of 8kpc in M33,
Williams et al 2009).

\section{Angular momentum transfers, radial migration}

There is a large efficiency in the exchange of angular momentum,
by resonant scattering at resonances, due to several successive
spiral patterns, as shown by Sellwood \& Binney (2002),
leading to radial migration of stars and gas (cf Figure 1).

\begin{figure}[b]
 \includegraphics[width=13cm]{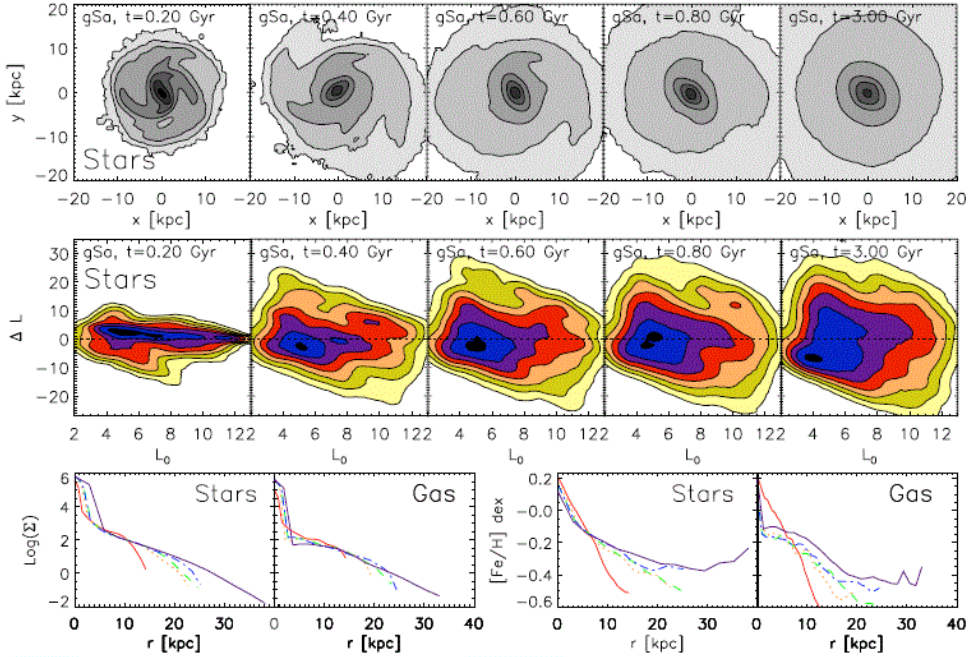} 
 \caption{Results of a Tree-SPH simulation, studying the exchange of angular momentum
due to resonance overlap of bar and spiral waves.
{\bf Top row:} Stellar disk density contours of a  giant Sa galaxy in 5 snapshots.
{\bf Middle  row:} Changes in angular momentum, $\Delta$L as a function of the 
initial angular momentum, L$_0$.  The locations of the
bar CR and OLR are indicated by the dotted and solid lines, respectively. 
{\bf Bottom row:} The evolution of the radial profiles of surface density (left) and
metallicity (right) for the stellar and gaseous disks. 
The initial disk scale-lengths are indicated by the solid lines. The 5 time steps
shown are as in the Top row, indicated by solid red, dotted orange, dashed green, 
dotted-dash blue and solid purple, respectively,
from Minchev et al (2010).}
   \label{fig2}
\end{figure}

The principle is that, at corotation, it is possible to exchange
angular momentum $L$ almost without heating. Assuming a nearly
steady spiral wave, at least for a few rotations, there is an energy invariant,
the Jacobi integral in the $\Omega_p$ rotating frame: 

$E_J = E  - \Omega_p L$, 

and the energy and angular momentum
are related by  $\Delta E  = \Omega_p \Delta L$. To separate energy
in the radial motions, defining the radial action $J_R$, it can be shown that:

  $\Delta J_R  = \frac{\Omega_p - \Omega}{\kappa} \Delta L$. 

This shows that at corotation, where $\Omega_p = \Omega$,
changes in $L$ do not cause changes in $J_R$, i.e. no radial
heating.
  Exchanges of $L$ will be most efficient near CR, and
the orbits which are almost circular will be preferentially scattered,
as shown in Figure 1.
 This kind of change in $L$ without heating is called 
churning by Sellwood \& Binney (2002), while 
the change with increase of epicyclic amplitude, through heating,
is called blurring, since it has specific signatures in the stellar orbits.
Gas contributes to churning, while it is also radially driven inwards.

In real galaxies, spiral waves are not steady, but transient,
and develop with different pattern speeds, so the CR could
span a wide range in radii. Radial migration could then involve
most of the galaxy disk.

\begin{figure}[b]
\begin{center}
 \includegraphics[width=8.2cm]{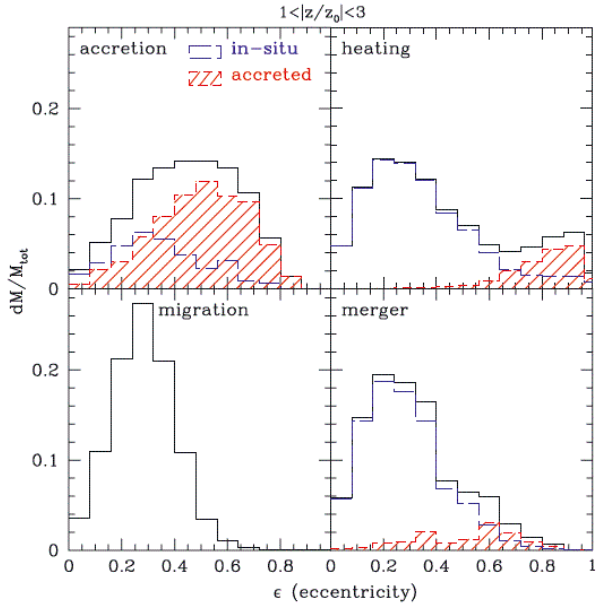} 
 \caption{Can the eccentricity distribution of stellar orbits
discriminate between the formation scenarios of the thick disk? 
Here the distributions are compared between four thick disc
formation models (accretion, heating, migration, merger), 
for stars in the range 1.3 scaleheights and
cylindrical distance $2<R/R_d < 3$,
 from Sales et al (2009).}
   \label{fig3}
\end{center}
\end{figure}

This has important consequences on
the chemical evolution, and abundance gradients.
The relation between metallicity and age is 
considerably scattered around a gross trend, as is observed
(Shoenrich \& Binney 2009). There is also a large
scatter in the  O/Fe versus Fe/H relation.
 Radial migration can also produce a thick disk,
since stars in the inner disks have higher z-velocity dispersion,
due to the higher restoring force of the disk. When migrating 
outwards, they will feel a smaller restoring
force towards the disk, and they will flare. This could explain both the
$\alpha$-enriched and low metallicity property of the thick disk.

The transfer of angular momentum can be 
multiplied if several patterns exist, with resonances in common.
Both bars and spirals can 
participate to this overlap of resonances.
Minchev \& Famaey (2010) have shown that due to the non-linear coupling,
the region affected by the migration is widened, and the 
migration rate accelerated  by a factor 3.
Numerical simulations have confirmed the high migration
rate, due to coupled patterns (Minchev et al 2010).
The exchange of angular momemtum has now maxima not only
at corotation, but also at the OLR.
The presence of the gas increases the rate of angular momentum exchange 
by about 20\%. Metallicity gradients can flatten in less than 1 Gyr 
as shown in Figure 2.
Again, the migration is the most important, for almost circular orbits.
It can explain the absence of clear age metallicity relation, or
age velocity relation.

\section{Bulge formation}

Several mechanisms are invoked to form bulges or spheroids.
First, as suggested by Toomre \& Toomre (1972), major mergers
of spiral galaxies can form an elliptical, with a remaining
disk, according to the relative alignment of their initial 
spins (e.g. Bendo \& Barnes, 2000). 
In minor mergers, disks are more easily kept and progressively enrich the
classical bulge.
Secular evolution can also form spheroids:
 bars and their vertical resonance elevate stars in the center
into what is called a pseudo-bulge. 
The latter is intermediate between a spheroid and a disk.
Pseudo-bulges are observed more frequently in late-type galaxies.
A third mechanism is provided by
clumpy galaxies at high redshift. The massive clumps
through dynamical friction, quickly spiral in, and form a bulge.
 Bulges in fact are too easily formed, and the main problem
of the hierarchical scenario is to form bulgeless galaxies.
These are observed with unexpected abundance, at z=0.

As for mergers, it is possible that most ellipticals
are the result of multiple minor mergers, instead of a 1:1
mass ratio remnant. Numerical simulations have shown that
the issue is not the mass ratio of individual mergers, 
but the total mass accreted, if at least
30-40\% of initial mass.
A large number of successive minor mergers can form an
elliptical galaxy, for instance 50 mergers
of 50:1 mass ratio. Given the mass function of galaxies,
multiple minor mergers are
even more frequent than equal-mass major mergers
(Bournaud, Jog Combes 2007a).

The formation of a box-peanut bulge, from
disk stars which are in vertical resonance with the bar,
i.e. when $\Omega - \nu_z/2 = \Omega_b$, where
$\nu_z$ is the vertical oscillation frequency,
has been known for a long time 
(Combes \& Sanders 1981, Combes et al 1990).
During secular evolution, the resonance can move
radially, following the slowing down of the bar,
and several peanuts can be formed, larger and larger
(Martinez-Valpuesta et al 2006).
Even after the bar destruction, the central stellar 
component will keep its thickness, forming a
pseudo-bulge. These spheroids, formed in numerical
simulations, correspond perfectly to the kind of
pseudo-bulges so frequently observed in late-type
galaxies.
They have characteristics intermediate between a classical 
bulge (or an elliptical) and a normal disk (Kormendy \& Kennicutt 2004),
their main properties are:
\begin{itemize}
\item
a luminosity distribution $\mu$ (in magnitude),
with a Sersic index $\mu \sim r^{1/n}$, with $n =1-2$  (exponential
disks have $n=1$, elliptical galaxies $n=4$ or larger)
\item
a flattening degree similar to disks, with
typical box/peanut shapes, and blue colors
\item
their kinematics show more rotation support than classical bulges.
\end{itemize}

Concerning the early evolution of galaxies, their percentage of gas is so high
that their disk is unstable and fragments in a few massive clumps.
With stars forming in those clumps, this scenario
explains the formation of the clumpy galaxies
observed at high redshift, which appear as
chain galaxies, when edge-on. Numerical simulations then show
 the rapid formation of exponential disks
and bulges, through dynamical friction
(Noguchi 1999, Bournaud et al 2007b).
The evolution time-scale might even be shorter than 
with spirals and bars, given the contrast of the structures.

\bigskip

Many spiral galaxies are observed with a thick disk,
distinct from their younger thin disk, where
contemporary stars are forming. This thick disk could 
trace some features of the galaxy past formation.
 At least 4 scenarios have been invoked for the
thick disk formation: 
\begin{itemize}
\item
1) Accretion and disruption of satellites (like in the stellar halo)
\item
2) Disk heating due to minor merger
\item
3) Radial migration, via resonant scattering
\item
4) In-situ formation from thick gas disk (mergers, or clumpy galaxies).
\end{itemize}

Sales et al (2009) recently proposed that
orbit excentricity of stars could help to disentangle
among these scenarios.

\section{Conclusion}

The gravitational stability of disks, the dynamics of
spiral perturbations, and the feedback cycle generated may explain the formation 
of exponential disks, with efficient kinematic viscosity.
 Angular momentum exchange are also very efficient with bars,
which gravity torques drive the gas towards the center, in a few
dynamical times.
 In addition,  resonant scattering by spiral waves,
constantly re-generated with different pattern speed, are
efficient to produce radial migration of stars and gas.
When overlap of resonances occurs, between spirals, or
between spiral and bars, the migration is strong and 
spans the whole disk.

If classical bulges and spheroids are commonly the results of mergers,
secular evolution can also form the pseudo-bulges, through
vertical resonance with the bar. Bulges in early and late-type galaxies
could be formed by a combination of these two mechanisms.
 Presently, it is very difficult to explain the presence of a 
large number of bulge-less galaxies, the more so as 
clumpy galaxies formed by gas instabilities at high redshift
will also form massive bulges through dynamical friction.

Several scenarios have been invoked for thick disk formation,
and a detailed study of stellar orbits could help to discriminate
among them.

\begin{discussion}

\discuss{J. Toomre}{There are large amounts of gas required for accretion. Is there evidence of this
infalling gas, and how could this gas accretion be tested, or observed around galaxies?
}

\discuss{F. Combes}{It is difficult to find direct evidence, since the infalling gas is diffuse.
Around our own galaxy, high velocity HI clouds have been observed for a long time,
and with simple models giving their distance, they would correspond to an infall
of a few solar masses per year.  Around external galaxies, the search is diffcult, by
lack of sensitivity. It is possible that the ubiquitous warps in spiral galaxies
come from external accretion. A stream of HI gas emission has been observed in the edge-on
galaxy NGC 891 up to 25kpc distance. It is interesting to note that cosmological
simulations predict gas accretion with the right order of magnitude, to maintain 
spiral and bars in galaxies.
}

\end{discussion}

\end{document}